\documentclass[%
 reprint,
 amsmath,amssymb,
 aps,
]{revtex4-2}
\usepackage{mathrsfs}
\usepackage{graphicx}
\usepackage{dcolumn}
\usepackage{bm}

\begin{document}

\preprint{APS/123-QED}

\title{Analytical Scaling Laws for Radiofrequency Based Pulse Compression \\ in Ultrafast Electron Diffraction Beamlines}

\author{Paul Denham}%
 \email{pdenham@physics.ucla.edu}
\author{Pietro Musumeci}
\affiliation{%
Department of Physics and Astronomy \\
University of California at Los Angeles, Los Angeles, CA, 90095}%

\date{\today}

\begin{abstract}
In this paper, we use an envelope equation-based approach to obtain analytical scaling laws for the shortest pulse length achievable using radiofrequency (RF) based bunch compression. The derived formulas elucidate the dependencies on the electron beam energy and beam charge and reveal how to obtain bunches containing 1 million electrons with single-digit fs pulse lengths; relativistic energies are strongly desirable. The effect of the non-linearities associated with the RF curvature and the beam propagation in a drift is shown to strongly limit the attainability of extreme compression ratios. Using an additional higher frequency RF cavity is proposed as a method to linearize the bunch compression and enable the generation of ultrashort beams in the sub-femtosecond regime. 
\end{abstract}

\maketitle


\section{\label{sec:intro}Introduction}
Ultrafast electron scattering requires the generation of very short electron bunches to capture the fastest physical processes \cite{zewail, sciainiMiller}. Due to the repulsive effect of space-charge forces, one critical challenge in this field is related to packing as many electrons as possible in a short bunch \cite{siwick2003}. In ultrafast electron diffraction (UED), pushing the electron energy to relativistic levels has helped in minimizing the space-charge effects, while at the same time bringing other advantages such as longer penetration depths, reduced group velocity mismatch, and suppressed inelastic scattering background \cite{weathersby2015mega, zhu2015femtosecond, manz2015mapping, filippetto2016design}. Over recent years, UED beamlines have seen continuous improvement in the achievable temporal resolution thanks to the introduction of techniques borrowed from accelerator physics based on the use of time-dependent radiofrequency (RF) electric field to compress the electron bunch during its propagation in the beamline  \cite{van2007electron}.  RF compression using 3 GHz resonant cavities has been applied to both non-relativistic and relativistic electron beamlines for UED \cite{van2010compression, maxson2017, zhao2018terahertz}, yielding bunch lengths down to the single-digit fs in the latter case. 

As we focus the discussion on the electron bunch length, one should also recognize that there are many factors other than the temporal duration of the probe pulse that contribute to the actual temporal resolution limit in a specific UED setup such as temporal jitter, group velocity mismatch, laser pulse length. For example, to counteract the additional temporal jitter introduced by RF-based compression, naturally synchronized laser-generated higher frequency waves have been used to impart an energy chirp on the beam in more complex coupling structures and drive the compression dynamics \cite{snively2020femtosecond, kealhofer2016all}.

In any case, though, to push the boundary of the UED technique, it is important to fully understand the limits in beam compression and how the various beamline parameters such as charge, energy, cavity voltage, and frequency affect the shortest bunch duration achievable. The minimum bunch length at the sample is the result of a complex interplay between the details of the bunching dynamics and the longitudinal space-charge forces in the beam so that typically UED practitioners have resolved to particle tracking simulation codes to design the beamline and predict the beam dynamics. The agreement with experimental results has been excellent \cite{GPT}, but particle simulations only deal with specific beamline setups, typically lack generality, and might not offer an immediate answer to the question of how to improve the compression in a given configuration.

It would be extremely useful to have a unified formalism for the treatment of the beam dynamics in RF-compression UED beamlines. This is the scope of this paper, which makes use of the longitudinal envelope equation formalism to include both the effects of longitudinal emittance and space-charge forces on the pulse evolution. The important contribution of the non-linearities in the beam compression is taken into account as an additional term in the longitudinal beam emittance. The result is an elegant analytical formulation for the minimum bunch length at the longitudinal waist which can be used to guide the simulation optimization, to compare parameter choices at different facilities, and more importantly to evaluate mechanisms for further improving the bunch length.


\begin{figure*}[ht]
    \includegraphics[scale=0.75]{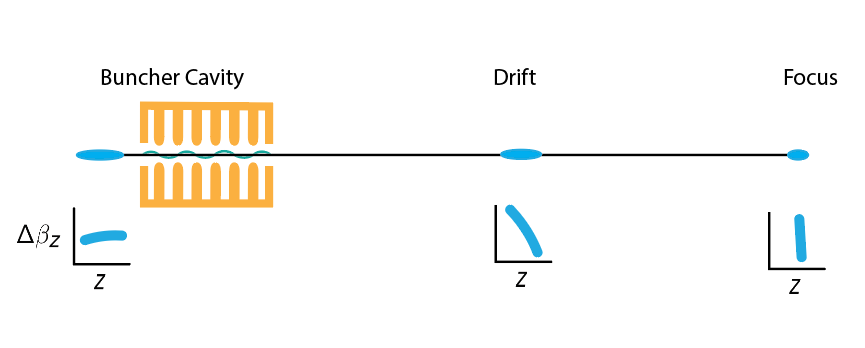}
    \caption{Illustration of RF ballistic bunching scheme. A velocity chirp is imparted on an electron using an RF cavity so that the tail of the beam has higher energy than the head. During the following drift the particles in the tail catch up with the particles in the head resulting in strong longitudinal compression} 
    \label{fig:bunching_scheme}
\end{figure*}

The simple cartoon in Fig. \ref{fig:bunching_scheme} is used to illustrate the dynamics under study. Essentially, a finite length electron beam is set to arrive inside an RF buncher cavity where electromagnetic fields oscillate with angular frequency $\omega = k c$ at the zero-crossing phase. Ideally, the input bunch length satisfies $k \sigma_z<<1$ so only a small phase window of the wave is sampled by the beam and the chirp imparted on the beam is predominantly linear. However, in our discussion, we will keep the higher-order terms in the energy modulation expansion to elucidate the role they play in the final bunch length. In the propagation region after the buncher, due to the strong energy chirp, the tail of the beam begins to catch up, while the head of the beam slows down. At some location downstream of the buncher, ideally arranged to be the sample plane or the interaction point of the UED experiment, the minimum bunch length is reached when the phase space distribution is vertically aligned. 

\begin{table}[hb]
   \centering
   \caption{Simulation beam parameters}
   \begin{tabular}{lcc}
       \toprule
       \textbf{Parameter} 
       &
       \textbf{High Energy}
       &
       \textbf{Low Energy}
       \\
       \hline
Focal length & 1.88 m & 1 m\\
Beam kinetic energy & 4.6 MeV & 150 keV \\
Norm. transverse emittance & 100 nm  & 8.3 nm \\
RMS transverse beam size & 100 um & 100um \\
Cavity Frequency & 2.856 GHz &2.856 GHz  \\ 
Relative energy spread &$10^{-5}$ & $10^{-5}$ \\
\hline
   \end{tabular}
   \label{parameters}
\end{table}

We will strive to keep all the formulas in the paper as general as possible (for example, not assuming $\beta = 1$) so that they could be applied to general RF compression setups (non-relativistic, MeV UED beamlines, as well as higher frequency compression schemes) once the parameters are scaled accordingly. For this reason, we will use two different example cases, loosely based on the UED beamlines at the UCLA Pegasus laboratory \cite{maxson2017}, to benchmark the agreement between the analytical framework and particle tracking simulations. The reference parameters used for this study are reported in Table \ref{parameters}.  


The main goal of this work is to provide simple and general expressions that can be used to evaluate the minimum achievable bunch length in a given beamline. In practice, we will first introduce the envelope equation formalism to describe the evolution of the bunch duration through the system. In the approximation that the buncher can be represented as a thin lens, we will then compute the leading contribution to emittance growth and analyze the ballistic dynamics in the drift in the absence of space-charge. We will then include the space-charge repulsion term in the envelope equation first assuming a Gaussian temporal profile for the beam and show how this sets the ultimate limit to the minimum achievable bunch length. Finally, we will expand the formalism to describe different current profiles and discuss the possibility of non-linearities compensation through the use of an additional higher frequency RF cavity to reach sub-fs bunch lengths as first proposed in \citet{floettmann:subfs}

\section{Longitudinal envelope equation}
\subsection{Envelope equation}
Among many possible choices for defining bunch duration (full width half maximum FWHM, full width containing 50 $\%$ of the charge FW50, etc.), in our discussion, we select the second-order moment of the longitudinal profile or root mean square (RMS) bunch length as the main quantity to follow the evolution of in the RF compression beamline. This is defined as:
\begin{equation}
    \sigma_z=\sqrt{\langle z^2\rangle}
\end{equation}
where $z$ is the longitudinal particle coordinate relative to the center of the bunch. $\langle ... \rangle$ represents an expectation value over the beam distribution function $f(z,z')$, and $z'$ is the relative velocity deviation from the average beam velocity. All the other definitions (FWHM, FW50, etc.) are simply proportional to the RMS bunch length using a proper order-of-unity pre-factor which depends on the beam distribution. The RMS beam size evolves along the beamline coordinate $s$ according to the equation \cite{reiserbook}:
\begin{equation}
\sigma_z''=\frac{\langle zz''\rangle}{\sigma_z}+\frac{\epsilon_{z,z'}^2}{\sigma_z^3}
\label{Eq:envelope}
\end{equation}
where the quantity $\epsilon_{z,z'}^2$ is the square of the RMS longitudinal trace space emittance, explicitly written in terms of the distribution moments as:
\begin{equation}
    \epsilon _{z,z'}^{2}=\langle z^2 \rangle \langle z'^2 \rangle -\langle z z' \rangle ^{2}
\end{equation}
It is well known that this quantity is conserved during transport if the forces acting on the particles are linear in the $(z,z')$ coordinates \cite{wiedemann}. 

Ideally, the longitudinal equation of motion is uncoupled from the transverse coordinates. On the other hand, the force acting on the particles might have a transverse dependence (i.e. $z''=\mathcal{F}(z,r,s)$) and in general the first term on the right side of Eq. \ref{Eq:envelope} will be proportional to $\langle z\mathcal{F}(z,r,s)\rangle$. Heuristically, we separate the contributions to the longitudinal force into a term associated with the external RF fields and a term proportional to the beam current which encodes the effect of the space-charge force i.e. $\mathcal{F} = \mathcal{F}_{RF} + \mathcal{F}_{sc}$. The latter, after averaging over the beam transverse distribution, will bear a dependency on the RMS transverse size $\sigma_r$, effectively coupling the longitudinal and transverse envelope equations. 

In most practical cases we can model the RF buncher as a thin lens and $\mathcal{F}_{RF}(z)$ simply provides an impulse force proportional to $\delta(s)$. In this case, the only effect of the RF buncher is to modify the initial conditions for the envelope equation through the application of a negative energy modulation to the phase space which causes the beam to begin the bunching process (i.e. $\sigma_{z0}'<0$). If the applied energy chirp is non-linear (which is the most common case), particular care has to be applied in evaluating the longitudinal emittance and the initial conditions at the exit of the buncher.

Meanwhile, the space-charge term $\langle z\mathcal{F}_{sc}\rangle(s,\sigma_z,\sigma_r)$  will act over the whole time of flight in the drift following the buncher.  To first order, this term is related to the Taylor expansion of the space-charge-induced longitudinal electric field near the center of the bunch. We will discuss this in detail at a later section of the paper, but to orient this initial discussion we can anticipate some of the results derived below. Typically this term can be evaluated analytically for simple longitudinal distributions and simply written as
\begin{equation}
    \frac{\langle z\mathcal{F}_{sc}\rangle(s,\sigma_z,\sigma_r)}{\sigma_z} = \frac{K_L}{\sigma_z^2}
\end{equation}
where $K_l \propto \tilde{g} N r_c / \beta^2 \gamma^5$ is the longitudinal perveance, which is comprised of the classical electron radius, $r_c$, the number of electrons in the bunch $N$, and $\tilde{g}$ is a geometry factor which depends on the beam aspect ratio. 

In what follows, we will assume a constant transverse RMS spot size for the beam and simplify the $s$ and $\sigma_r$ dependencies in $K_l$, which allows solving the longitudinal envelope equation independently from the transverse dynamics. Somewhat surprisingly, this crude approximation works fairly well since if the initial bunch length in the beam rest frame is sufficiently long, the space-charge term is only weakly affecting the dynamics and can be neglected. However, the space-charge effects eventually begin to matter as we close in on the optimal bunching conditions. In addition, it is important to note that whenever the charge density is sufficiently high, the non-linearities in the bunch self-fields will cause emittance growth during the propagation. This is the situation where our approach of solving the envelope equation assuming a constant longitudinal trace space emittance falters. In this case, one will need to resort to a kinetic approach where each particle trajectory is self-consistently tracked in the field generated by the charge distribution. Nevertheless, this situation should be avoided at all costs if ultrashort bunch lengths are desired, and the effects of space-charge must be kept relatively small. Therefore the analytical model where only the first-order linear space-charge contribution is taken into account does provide an excellent approximation for the shortest bunch lengths achievable with RF compression.

\subsection{Solution in a drift}
Let us start the discussion from the simplest and physically relevant case in which the propagation occurs in a drift and space-charge forces can be neglected, i.e. $\mathcal{F}=0$. 
The shortest bunch length along the line can then be found recasting Eq. \ref{Eq:envelope} as:
\begin{equation}
\frac{1}{2}\frac{d}{ds} \left(  \sigma _{z}'^{2} \right)  =  \frac{\epsilon_{z,z'}^{2}}{\sigma_z^{3}}  \sigma_z'
\end{equation}
which can be integrated exactly:
\begin{equation}
    \sigma_{zf}'^2-\sigma_{z0}'^2=\epsilon_{z,z'}^2\left(\frac{1}{\sigma_{z0}^2}-\frac{1}{\sigma_{zf}^2}\right)
\end{equation}
The waist position is a local minimum for $\sigma_z$, thus we can set $\sigma_{zf}'=0$. If the RF buncher imparts an initial condition $\sigma_{z0}'\approx-\frac{\sigma_{z0}}{f}$ so that the longitudinal waist is reached after a propagation distance $f$, we have
\begin{equation}
    \sigma_{zf}=\frac{1}{ \sqrt{\frac{1}{\sigma_{z0}^2}+\frac{\sigma_{z0}^2}{\epsilon_{zz'}^2 f^2}}}\approx\frac{f \epsilon_{z,z'}}{\sigma_{z0}}
\label{Eq:min_bunch_length}
\end{equation}
where we neglected the term $\frac{1}{\sigma_{z0}^2}$ which for large compression factors is always much smaller than $\frac{1}{\sigma_{z}^2}$.
This equation simply says that beams with smaller longitudinal emittance enable reaching shorter bunch lengths. It also follows that by decreasing the focal length $f$ of the RF buncher, proportionally shorter final bunch durations can be achieved.

\begin{figure*}[ht]
    \includegraphics[scale=0.55]{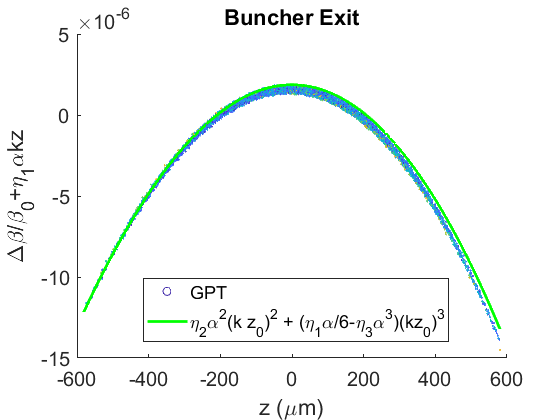}(a)
    \includegraphics[scale=0.55]{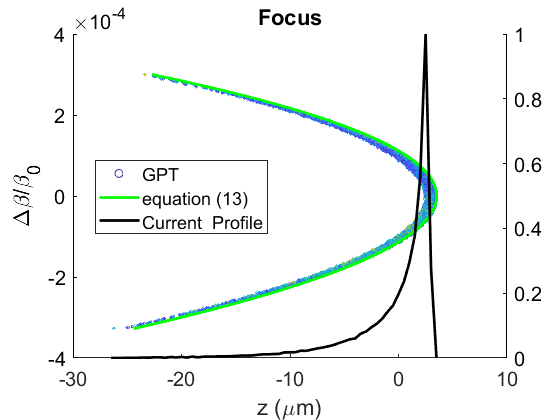}(b)
    \includegraphics[scale=0.55]{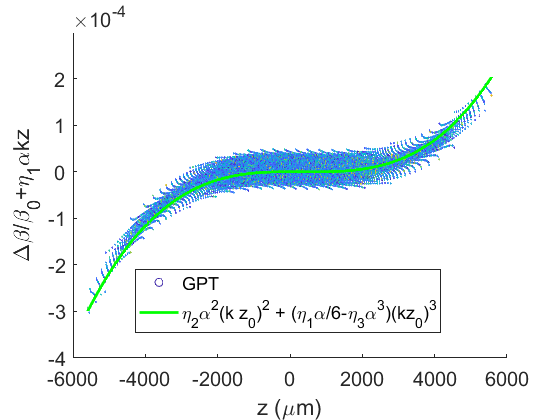}(c)
    \includegraphics[scale=0.55]{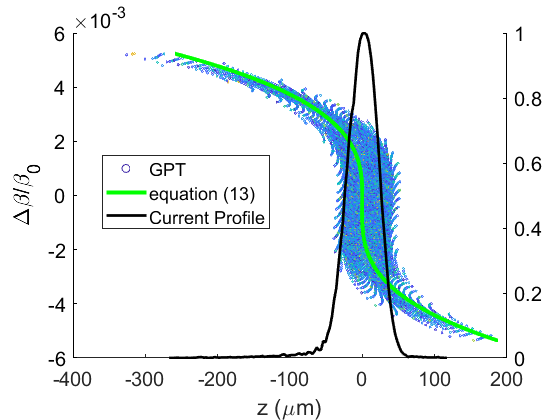}(d)
    \caption{Left) Trace spaces of the beam at the exit of the prebuncher after the linear chirp has been subtracted from the distribution for high energy (top) and low energy (bottom) cases, compared with the analytical predictions from Eq. \ref{Eq:buncher_exit} (green curves). Right) Longitudinal trace spaces at the temporal waist for the high energy (top) and low energy (bottom) cases compared with the predictions from Eq. \ref{Eq:finalphasespace} (green curves). The current profiles at the focus are also shown in black.}
    \label{fig:phase_space_check}
\end{figure*}

\subsection{Single particle dynamics and non-linear phase-space correlations in the RF buncher}
Let us now look more closely at the details of the negative energy chirp imparted by the RF buncher on the beam distribution. The main assumption here will be that the cavity fields act on the electrons by adding a sinusoidal energy chirp which depends on the particle position within the bunch:
\begin{equation}
    \Delta\gamma=-\alpha\sin(kz)
\end{equation}
where $\alpha=eV_0/mc^2$ and $eV_0$ is the cavity voltage or maximum energy gain seen by an ideally phased particle, and $k=k_0/\beta$ is the RF angular wave number divided by the normalized longitudinal velocity. The phase of the cavity is tuned so that the center of the bunch experiences no net energy gain and a predominantly linear energy chirp is imprinted on the bunch. Particles at the tail gain energy, while particles at the head of the bunch lose energy resulting in the compression dynamics in the following drift. 

There are two distinct sources of non linearities in the trace space dynamics resulting from the applied energy change to the particles. Firstly, for finite duration input bunches, the curvature of the RF wave will cause significant non linear effects in the trace space. In addition, the relativistic relation between normalized velocity and beam energy $\beta = \sqrt{1-1/\gamma^2}$ adds an important degree of non linearity to the transport as pointed out in \citet{zeitler}.  Taylor-expanding the relative velocity deviation $\frac{\Delta\beta}{\beta}$ in terms of the energy deviation we can write \begin{equation}
    \frac{\Delta \beta}{\beta} =\sum_m \eta_m\Delta\gamma^m
    \label{Eq:Deltabeta}
\end{equation} where $\eta_m$ are proportional to the mth-derivatives $\frac{d^m}{d \gamma^m}\beta$ and in particular
\begin{subequations}
\begin{align}
    \eta_1 &= \frac{1}{\beta^2\gamma^3} \\
    \eta_2 &= \frac{2-3\gamma^2}{2\gamma^6\beta^4} \\
    \eta_3 &= \frac{2-5\gamma^2+4\gamma^4}{2\gamma^9\beta^6}
\end{align}
\end{subequations}
where $\gamma$ and $\beta$ are the mean values of the normalized energy and velocity distributions respectively. The coefficients $\eta_m$ scale as $\gamma^{-(m+2)}$ so that at high relativistic energies the higher order non linear terms in the transport can be neglected. 

If we are interested in the lowest order dynamics, we can simply replace $\sin(kz)$ with $kz - \frac{(kz)^3}{6}$, and truncate the series after the first three terms to obtain the correlation of the velocity deviation with the initial particle position which can be written as: 
\begin{equation}
    \frac{\Delta\beta}{\beta} \approx -\eta_1\alpha (k z_0)+\eta_2\alpha^2(k z_0)^2 + \left(\frac{\eta_1\alpha}{6} - \eta_3\alpha^3 \right)(kz_0)^3
    \label{Eq:buncher_exit}
\end{equation}
where $z_0$ is the input particle longitudinal coordinate.
We verify this expression at high energy (4.6 MeV) and low energy (150 keV) by considering a particle tracking simulation of the buncher configuration listed in Table \ref{parameters} with an initial bunch length of 195~$\mu$m and 1.87~mm respectively. The buncher was modeled by a standing wave cylindrically symmetric TM010 cavity with an amplitude adjusted to reach a longitudinal focus 1.88m and 1m downstream respectively for the high and low energy cases. Since the cavity length is less than 0.05 m, it is reasonable to approximate it as a thin lens. The longitudinal phase spaces from GPT at the exit of the buncher are shown in Fig. \ref{fig:phase_space_check}(a) and \ref{fig:phase_space_check}(c) for the high and low energy case respectively with subtracted linear correlations. The quality of the agreement between GPT and our analytical framework can be assessed by comparing the distributions with the lines corresponding to Eq. \ref{Eq:buncher_exit} which are also shown. The parameters chosen for these examples highlight the different possibilities for the dominant non-linearity in the system. In the high-energy case, the relativistic effects are responsible for the parabolic shape seen in the simulation. While in the low energy case, the injected bunch length is longer and the third-order non-linearity associated with the sinusoidal RF fields is the main effect in the beam distribution shape. 

The convenience of working in the trace space is the linearity of the dynamics in the drift which fully preserves the phase space area. Explicitly, in the drift after the buncher, the longitudinal particle position can be written as
\begin{equation}
    z=z_0+s\frac{\Delta\beta}{\beta}=z_0+s\sum_{n=1}^{\infty}\eta_n\Delta\gamma^n
    \label{Eq:driftpropagation}
\end{equation}

The initial coordinate $z_0$ is expressed in terms of the induced energy modulation $\Delta \gamma$ by inverting Eq. 8. Then $\Delta\gamma$ is Taylor expanded in terms of $\Delta\beta/\beta$. Substituting into Eq. \ref{Eq:driftpropagation}, keeping only terms up to third order, we can write:
\begin{multline}
    z \approx \left(s-\frac{1}{\eta_1\alpha k}\right)\frac{\Delta\beta}{\beta} +
    \\ 
    +\frac{\eta_2}{\eta_1^3\alpha k}\frac{\Delta\beta^2}{\beta^2} - \frac{\left(\frac{\eta_1 \alpha}{6}- (\eta_3-2\eta_2^2/\eta_1)\alpha^3  \right)}{\eta_1^4 \alpha^4 k} \frac{\Delta\beta^3}{\beta^3}
    \label{Eq:finalphasespace}
\end{multline}
The longitudinal waist occurs where the linear chirp is cancelled at distance $s=\frac{1}{\eta_1\alpha k}$ along the beamline allowing us to define the buncher longitudinal focal length:
\begin{equation}
    f=\frac{1}{\eta_1\alpha k} = \frac{m_0c^2\gamma^3\beta^2}{eV_0k}
    \label{Eq:focal_length}
\end{equation} 
which indicates that very high voltage cavities are needed to obtain short focal lengths with relativistic electrons. It is also useful to note the k-dependence of this expression which suggests the use of very high frequencies for this application. 

At the focal plane, the residual correlation is quadratic or cubic in $\Delta\beta$ depending on the relative importance of the non-linearity in the drift propagation concerning the RF curvature. As discussed above, lower beam energies and longer input bunches tend to show higher third-order non-linearities, while relativistic energies typically have dominant second-order contributions. The predictions from Eq. \ref{Eq:finalphasespace} can be again verified by comparing to the phase spaces at the temporal waist plane from the same GPT simulation, as shown in Fig. \ref{fig:phase_space_check}(b) and (d).

\subsection{Emittance growth mechanisms and the relationship between different longitudinal phase space definitions.}

\subsubsection{$(z,z')$ trace space emittance}
\label{sect:rfemittancegrowth}

Since the drift dynamics in the trace space are entirely linear, and the emittance growth is all accrued in the buncher, the envelope equation formalism is a convenient choice to follow the RMS bunch length evolution. To evaluate the RMS emittance growth induced by the RF compressor, we start from an initial longitudinal phase space with RMS emittance $\epsilon_{z_0,z_0'}$. After the energy chirp is applied, then the single-particle velocity variation maps as $z'_0\to z_0'+\Delta\beta/\beta$, where $\Delta\beta/\beta$ represents the velocity variation imparted by the buncher, which is correlated with particle position. In the thin lens approximation, the particles do not change position as the beam goes through the cavity.

\begin{figure}[ht]
    \includegraphics[scale=0.55]{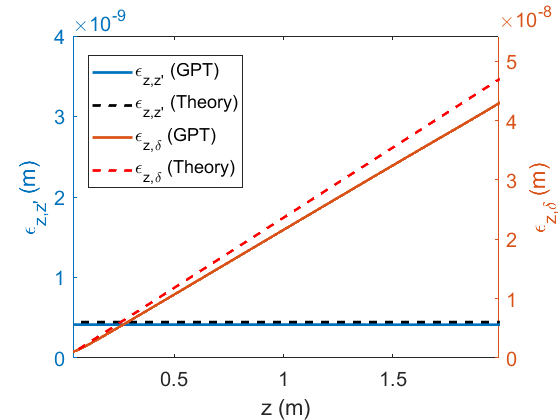}
    \caption{Comparison between analytical estimates (dashed lines) and GPT simulations (solid lines) for an initial bunch length of 0.65 ps. The trace space emittance after the buncher is shown in blue, and emittance growth in $(z,\delta)$ phase space  in orange.}
    \label{fig:emittance_growth}
\end{figure}

The moments of the new distribution can be calculated and the relation between initial and final emittance after the buncher written as 
\begin{align*}
\epsilon_{zz'}^2 &=\epsilon_{z_0z'_0}^2+\epsilon_{RF}^2 \\
&= \epsilon_{z_0z'_0}^2+\langle z_0^2 \rangle \langle \left(\frac{\Delta\beta}{\beta}\right)^2 \rangle -\langle z_0 \left(\frac{\Delta\beta}{\beta}\right)\rangle^{2}
\end{align*}
Note that each respective expectation value needed to calculate the emittance, with the exception of $\langle z_0^2\rangle = \sigma_{z0}^2$, requires integrating $\sin^m(kz)$ or $z\sin^m(kz)$ over the beam distribution. Assuming an initial gaussian current profile, these integrals have closed-form expression up to arbitrary order of m, but the leading contributions to the emittance growth can be obtained simply expanding $\sin(kz) \approx kz$ and using the second-order term in the expansion of $\Delta\beta$
The result is:
\begin{equation}
    \epsilon_{RF}^2\approx \sigma_{z0}^2 \left[2 \eta_2^2\alpha^4 k^4 \sigma_{z0}^4 + \frac{1}{6}\left(\eta_1\alpha - 6\eta_3\alpha^3 \right)^2 k^6 \sigma_{z0}^6 \right]
\label{Eq:RFemittance}
\end{equation}
In most cases, this expression is much larger than the initial longitudinal emittance because the buncher is fundamentally inducing a large velocity spread (with correspondingly large non-linear contributions) to achieve compression. As long as the space-charge forces are negligible in the drift, $\epsilon_{RF}$ is also equal to the final emittance and can be used to calculate the limit for the shortest bunch length at the focus in Eq. \ref{Eq:min_bunch_length}. The emittance growth in the buncher and its preservation in the drift can be seen in Fig. \ref{fig:emittance_growth}, where the trace space RMS emittance evolution calculated from GPT is plotted along the beamline. The final emittance after the short buncher cavity at the origin is well approximated by the analytical formula \ref{Eq:RFemittance} which is represented by the black dotted line. 

\subsubsection{$(z,\delta)$ phase space}
It is important to note at this point that if instead, we had utilized the more common choices of defining the longitudinal trace space in terms of the relative energy spread $\delta = \frac{\Delta \gamma}{\gamma}$ or momentum spread $\Delta p_z / p_z$, the drift dynamics would become highly non-linear especially for mildly relativistic particles. 

Initially, the trace space emittance $\epsilon_{z,z'}$ is related to the $(z,\delta)$ emittance by the following relationship:
\begin{equation}
    \epsilon_{z_0,\delta}=\gamma^2\beta^2\epsilon_{z_0,z_0'}
\end{equation}
This relationship holds because $z' \approx \eta_1 \Delta \gamma$ and explains the order of magnitude difference in absolute emittance values in Fig. \ref{fig:emittance_growth}. It is then fairly common to see in the literature the envelope equation written in terms of the $(z,\delta)$ emittance:
\begin{equation}
\sigma_z''=\frac{\langle zz''\rangle}{\sigma_z}+\frac{\epsilon_{z,\delta}^2}{\beta^2 \gamma^2 \sigma_z^3}
    \label{Eq:envelopezdelta}
\end{equation}
Nevertheless, in a drift, the particle positions evolve according to Eq. \ref{Eq:driftpropagation} where the transport is inherently non-linear, especially for mildly relativistic particles, causing emittance growth and limiting the usefulness of the envelope equation approach. Ultimately, for large enough initial energy spreads, the higher-order terms proportional to $\eta_m$ lead to significant $(z,\delta)$ emittance growth which can be estimated using the same techniques as in the previous section
\begin{equation}
    \epsilon_{z\delta}^2= \epsilon_{z_0,\delta_0}^2 + 2  s^2  \eta_2^2 \alpha^6 k^6\sigma_z^6
    \label{Eq:growthindrift}
\end{equation}
where $\epsilon_{z_0,\delta_0}$ is the emittance at the beginning of the drift. For a small initial emittance, this expression predicts a nearly linear growth with propagation distance. This is shown in Fig. \ref{fig:emittance_growth} where we also show the $(z,\delta)$ phase space evolution from GPT and compare it with Eq. \ref{Eq:growthindrift} with the inclusion of the initial emittance as well. 

The seemingly counterintuitive behavior of the longitudinal emittance (which grows linearly in the drift) is the main reason for us to adopt the $(z,z')$ trace space emittance in the calculation of the final bunch length when using the envelope equation formalism. Finally, for completeness, we observe that if the un-normalized momentum $(z,\Delta p_z/p_z)$ was used as a trace space variable, all expressions could be simply modified substituting $z' = \frac{1}{\gamma_0^2}\Delta p_z/p_z$. Nevertheless, due to the relativistic non-linear relation between momentum and velocity, even in this case, one would have significant emittance growth in the drift propagation. 

\subsection{Bunch length limit in absence of space-charge effects}

We are now in a position to discuss how the minimum achievable bunch duration scales with the main beamline parameters. Inserting the trace space longitudinal emittance estimate (Eq. \ref{Eq:growthindrift}) into the envelope-equation formalism (Eq. \ref{Eq:min_bunch_length}), we get an expression written as the quadrature sum of the different contributions to the final emittance between i) the initial uncorrelated relative energy spread $\sigma_\delta$, ii) the non-linearities introduced by the relativistic correction to the transport or iii) the RF-induced emittance.

Note that for practical reasons (related to the requirements of the sample chamber, pumping geometry, transverse optics), the focal length of an RF buncher is typically 1 m long or less in most setups. Therefore, in order to facilitate comparison between beamlines of different energies, we can rewrite the terms as a function of the focal length of the buncher. Assuming that one term in the sum is much larger than the others, we can synthesize this result as:
\begin{equation}
    \sigma_{zf} \approx \max 
    \begin{cases} 
    f \sigma_{z0'} = \frac{f}{\beta^2 \gamma^2}\sigma_\delta \\
    \noalign{\vskip9pt}
    \sqrt{2} \frac{|\eta_2|}{\eta_1}\alpha k \sigma_{z0}^2 \approx  \frac{3\sqrt{2}\gamma^2}{2} \frac{\sigma_{z0}^2}{f} \\
    \noalign{\vskip9pt}
    \frac{1}{\sqrt{6}}k^2\sigma_{z0}^3
    \label{Eq:simpleformulabunchlength}
    \end{cases}
\end{equation}
Note that the full quadrature sum of these three expressions should be used in those cases where two or more contributing terms have similar magnitude. 

In the absence of non-linearities in the transport, the final bunch length would be simply proportional to the initial relative velocity spread $\sigma_{z0}'$. Note that for the same uncorrelated relative energy spread $\sigma_\delta$, a MeV energy system has a clear advantage in this case in the ability to achieve ultrashort bunch lengths. 

At the same time, as the energy increases, it is most likely that the non-linearities due to the relativistic dynamics in the drift would become the dominant contribution to the final emittance and bunch length. In this regime, after approximating $\eta_2/\eta_1^2 \approx 3/2\gamma^2$ and use the definition of the focal length (Eq. \ref{Eq:focal_length}) the final bunch duration scales as the square of the beam energy. For longer initial bunch lengths, the curvature of the RF dominates the shape of the final phase space and sets the limit for the shortest bunch duration achievable. This contribution is important for both relativistic and non-relativistic energies and it strongly favors the use of lower RF frequencies. In all THz compression experiments carried out so far this term has been the main limit to the final bunch length. 

\begin{figure}[ht]
    \includegraphics[scale=0.55]{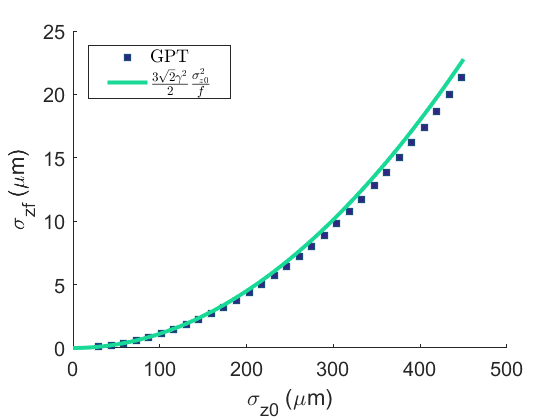}(a)
    \includegraphics[scale=0.55]{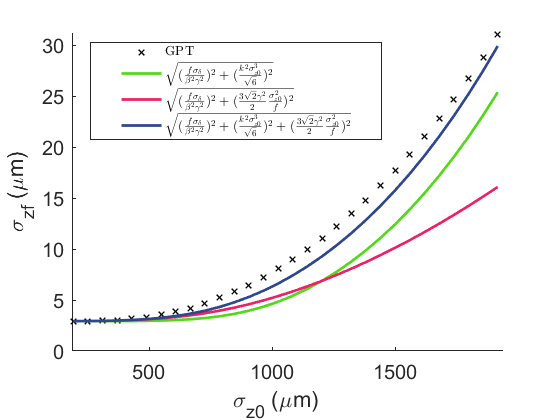}(b)
    \caption{Final bunch length as a function of the input bunch length for the case of (a) high energy, and (b) low energy. The analytical curves are also shown and are found to be in excellent agreement with the simulations.}
    \label{fig:waist0sc}
\end{figure}

The analytical formulas summarized in Eq. \ref{Eq:simpleformulabunchlength} were found in excellent agreement with GPT in both the relativistic and non-relativistic regimes. The results are shown in Fig. \ref{fig:waist0sc} top and bottom for the 4.6 MeV and 150 keV cases respectively. In the low energy case, the cubic non-linearity due to the RF curvature dominates for longer initial bunch lengths.

In the absence of space-charge, the shortest bunch durations are obtained by minimizing the emittance growth in the temporal lens which can be done by using very short input beams. At the same time, as we will see, decreasing the initial bunch length also increases the initial peak current so that the space-charge effects in the longitudinal envelope equation can no longer be neglected. At a certain point, space-charge effects begin to take over and prevent further bunch compression. Thus, there must be an optimum initial bunch length to inject which exactly balances the RF and space-charge emittance growth. We address this in the following section.

\section{space-charge limits to compression}

\begin{figure*}[ht]
    \includegraphics[scale=0.55]{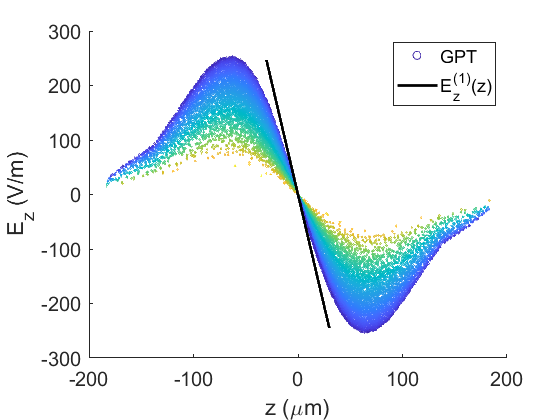}(a)
    \includegraphics[scale=0.55]{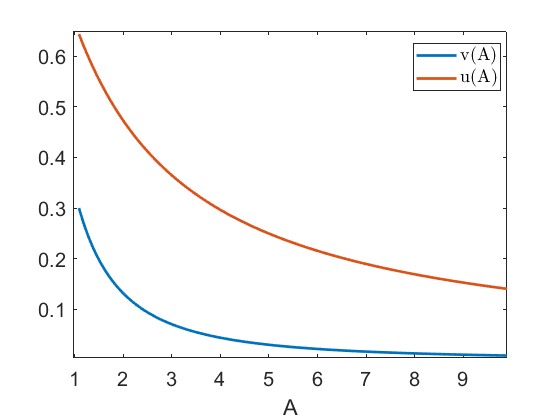}(b)
    \caption{(a) Longitudinal field of the bunch compared with the linear field component of the 3D Gaussian field. The total charge in the bunch in this simulation is $Q=10^5e$. (b) Geometry factors for the transverse (red) and longitudinal (blue) field plotted as a function of the rest frame aspect ratio.}
    \label{fig:gfactor}
\end{figure*}

\subsection{An example of geometry factor calculation: Gaussian distribution case}

To add in the envelope equation formalism the effect of the space-charge on the bunch length evolution, we need to compute a reasonable representation for the  $\langle z\mathcal{F}_{sc} \rangle(z,\sigma_z,\sigma_r)$ term. For a first-order estimate, it will suffice to retain the linear term in the longitudinal electric field felt by an on-axis particle in the bunch core. As space-charge effects become more important, the emittance growth caused by the higher-order terms in the profile of the electric field makes the envelope equation approach less useful for describing the bunch evolution and one would need to go back to self-consistent particle tracking simulations. To validate all of the estimates made in what follows, we simulate the bunching process utilizing GPT's spacecharge3Dmesh algorithm using the nominal beam parameters shown in Table \ref{parameters} and a beam charge of 16 fC (10$^5$ electrons).

For simplicity, we will start deriving the self-field of a cylindrically symmetric Gaussian beam assuming that the distribution function retains a Gaussian profile throughout the system. Admittedly this approximation is especially poor at the waist where the linear chirp is removed and what remains is a non-linear distribution in trace space with a characteristic current spike and a temporal profile strongly asymmetric and far from a regular Gaussian. Nonetheless, the linear force obtained below can at the very least be used to estimate the turning point in the system to understand when space-charge effects can no longer be neglected. The charge density in the beam rest-frame can be written as:
\begin{equation}
\rho=\frac{Q \exp \left(-\frac{r^2}{2 \sigma _r^2}-\frac{z^2}{2 \sigma _z^2}\right)}{(2\pi) ^{3/2} \sigma _r^2 \sigma_z}
\end{equation}
In order to obtain the linear components of the electric field, we Fourier transform the charge density and write the potential as \begin{equation}
    \phi(x,y,z)=\iiint\frac{\tilde{\rho}(k_x,k_y,k_z)}{\epsilon_0(k_x^2+k_y^2+k_z^2)}\exp(i\boldsymbol{k}\cdot\boldsymbol{r})\frac{d^3k}{(2\pi)^3}
\end{equation}
where $\tilde{\rho}$ is the Fourier transform of the charge density which has also gaussian shape. 

We can then expand the exponential to second order in $\left(\boldsymbol{k}\cdot\boldsymbol{r}\right)$. After dropping the constant term which is immaterial for the field profile, we also note that the first term vanishes by symmetry of $\tilde{\rho}$. The second order term yields the uncorrelated linear electric field components for a gaussian bunch distribution which, after performing a series a gaussian integrals, can be written as:
\begin{equation}
    \boldsymbol{E}^{(1)}=\frac{Q}{ (2 \pi )^{3/2}2\epsilon _0\sigma_z\sigma_r ^2}u(A)r\boldsymbol{\hat{r}}+\frac{Q}{ (2 \pi )^{3/2}\epsilon _0\sigma_z^3}v(A)z\boldsymbol{\hat{z}}
\end{equation}
where $A =\sigma_r/\sigma_z$ is the beam aspect ratio (in its rest frame). The predicted field gradients are found in very good agreement with core field gradients extracted from GPT simulation as shown in the example in Fig. \ref{fig:gfactor}(a). The geometry factors (plotted for reference in Fig. \ref{fig:gfactor}b) are given by:
\begin{equation}
u(A)=\frac{\xi(A)-(1-\xi(A)^2)\coth^{-1}\left(\frac{1}{\xi(A)}\right)}{\xi(A)^3}
\end{equation}
\begin{equation}
v(A)=\frac{\coth^{-1}\left(\frac{1}{\xi(A)}\right)-\xi(A)}{\xi(A)^3}
\end{equation}
where $\xi(A) = \sqrt{1-A^2}$.

To take advantage of this result, we must first express the longitudinal electric field just calculated in the beam rest frame, in terms of laboratory frame quantities, which can be done by rescaling the longitudinal coordinate and RMS moments by $\gamma$, i.e,  $E_z^{(1)}(z,\sigma_z,\sigma_r)\to E_z^{(1)}(\gamma z,\gamma \sigma_z,\sigma_r)$ .

The generalized force term is then determined noting that $z'' = \gamma'/\gamma^3\beta^2$ and then $\gamma'=eE_z^{(1)}/mc^2$. Putting all together, the space-charge term in the envelope equation can be written as:
\begin{equation}
    \frac{\langle z\mathcal{F}_{sc}(z,\sigma_z,\sigma_r)\rangle}{\sigma_z}=\frac{g(A) N r_c}{\beta^2 \gamma^5\sigma_z^2} = \frac{K_L}{\sigma_z^2}
    \label{Eq:longperveance}
\end{equation}
where $g(A)=\sqrt{2/\pi}v(A)$ is an order of unity factor which takes into account the geometry effects, $r_c$ is the classical electron radius and $N$ the number of electrons in the bunch and we have defined the longitudinal perveance $K_L$ as anticipated in the first section. 



As anticipated, the dependence of the geometry factor $g(A)$ on the beam aspect ratio requires to self consistently solve the transverse and longitudinal envelope equations as a coupled system. Since we are mainly interested in longitudinal dynamics, in this paper we will greatly simplify the transverse evolution and simply assume a slowly varying transverse RMS spot size along the beamline. To take into account the evolution of the aspect ratio during the compression, we are then left with two choices: i) either include in the longitudinal perveance term the entire dependence on the aspect ratio (which is evolving as the beam is compressing) and resort to numerical integration, ii) or pick a reasonable value for $g$ and assume it constant for most of the beamline which allows a direct integration for the bunch length evolution. This approximation turns out to be quite acceptable as for much of the bunching process, the space-charge effects are negligible and will only turn on near the waist when the beam density reaches a maximum. To this end, we can evaluate $g$ at the mean aspect ratio over the beam propagation distance neglecting space-charge i.e. 
\begin{equation}
    \bar{A}(\sigma_{z0})\approx\frac{1}{f}\int_0^f A(s)ds
    \label{Eq:meanaspectratio}
\end{equation}
where we highlight the dependence on the initial bunch length. We can then utilize $g(\bar{A})$ to integrate analytically the envelope equation. 



Assuming a transverse RMS waist $\sigma_r$ at the buncher and an RMS divergence $\sigma_{\theta}$, the aspect ratio as a function of drift distance can then be written as:
\begin{equation}
A(s)=\frac{\sqrt{\sigma_r^2+\sigma_{\theta}^2s^2}}{\gamma\sqrt{\sigma_{z0}^2+2s\sigma_{zz'}+s^2\sigma_{z'}^2}}
\label{Eq.ratio}
\end{equation}
where $\sigma_{z0}^2$, $\sigma_{zz'}$ and $\sigma_{z'}^2$ are the longitudinal beam matrix element at the exit of the buncher.
\begin{figure}[ht]
    \includegraphics[scale=0.55]{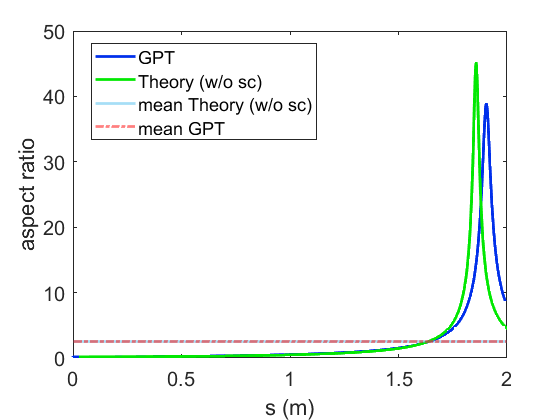}
    \caption{Comparison of aspect ratio from theory and GPT.}
    \label{fig:aspect}
\end{figure}
Although the space-charge fields are going to ultimately change the mean value of the aspect ratio, the difference is not significant when space-charge is only mildly influencing the dynamics (as we will see this is the case close to the optimum), as shown in Fig. \ref{fig:aspect}, where Eq. \ref{Eq.ratio} is compared with GPT using an initial bunch length of $67.5 \mu m$. There is some disagreement between the location of the peak and the maximum value due to the space-charge, nonetheless, the mean values are within $3 \%$ of each other.

\subsection{Effect of the longitudinal space-charge force on the minimum bunch length}

\begin{figure*}[ht]
    \includegraphics[scale=0.55]{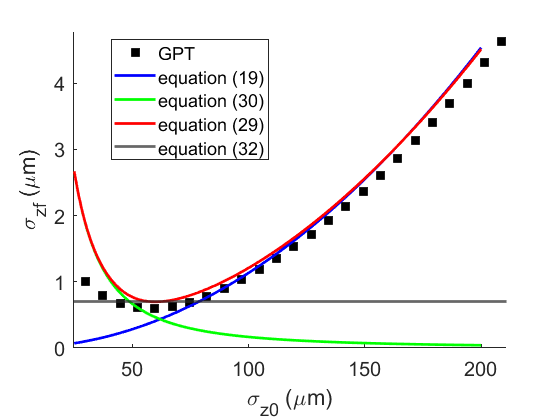}(a)
    \includegraphics[scale=0.55]{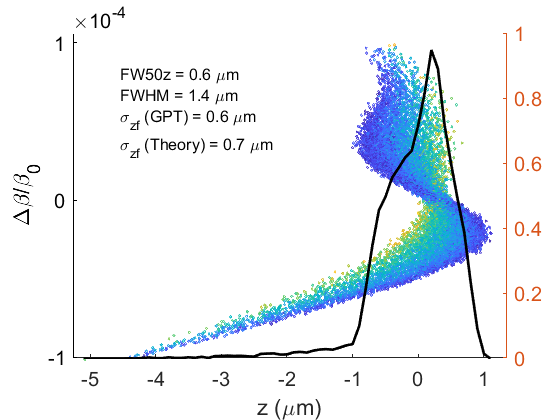}(b)
    \includegraphics[scale=0.55]{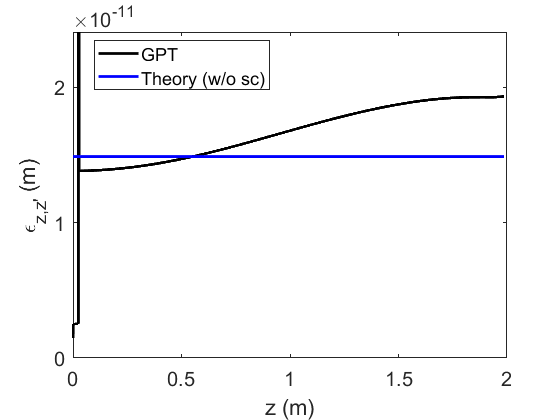}(c)
    \includegraphics[scale=0.55]{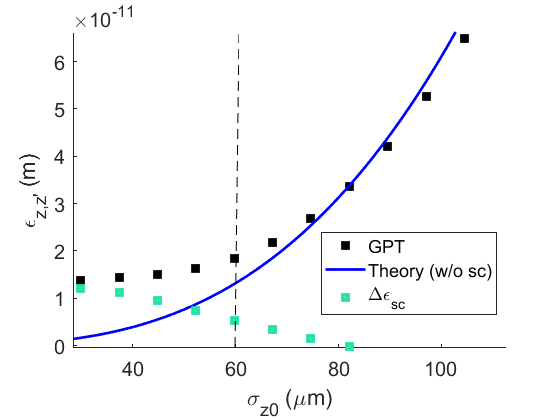}(d)
    \caption{(a) Final waist size plotted with respect to initial bunch length compared with GPT. The green and blue lines show the asymptotic behavior in the regimes where space-charge and emittance growth respectively are dominating the dynamics. The red line is the analytical expression which takes into account all effects discussed in this paper (b) The optimum phase space found from the scan in (a) for an input bunch length of 67.5 $\mu$m. Various measures of the bunch length are reported. The color-coding of the particles is a function of the radial coordinate (blue corresponds to on-axis particles). c) Evolution of the emittance along the line. The growth observed in GPT is due to the non-linearities of the space-charge field. d) Final emittance in GPT as a function of initial bunch length. The relative importance of the space-charge contribution to the emittance can be inferred by subtracting the expected emittance growth due to the buncher dynamics non-linearities. The cross-over point (where space-charge becomes the dominant effect) can be used to estimate the optimal injection condition and hence the minimum final bunch length achievable for a given setup.}
    \label{fig:scminagreement}
\end{figure*}

Using these results, the longitudinal envelope equation in presence of space-charge can be written as:
\begin{equation}
    \sigma_z''=\frac{K_L}{\sigma_z^2}+\frac{\epsilon_{z,z'}^2}{\sigma_z^3}
\end{equation}
where $K_L$ is the longitudinal perveance as defined in Eq. \ref{Eq:longperveance}. 

Following the same steps that led to Eq. \ref{Eq:min_bunch_length} we can directly integrate to calculate the bunch length at the waist, obtaining:
\begin{equation}
    \sigma_{zf}=\frac{\epsilon_{z,z'}^2}{\sqrt{ \frac{\sigma_{z0}^2 
    \epsilon_{z,z'}^2}{f^2}+K_L^2}-K_L}
\label{Eq:sigmazfwithsc}
\end{equation}
where we have once again assumed that the initial bunch length is much larger than the final one.

In the limit of $K_L$ very small, Eq. \ref{Eq:sigmazfwithsc} yields back the zero space-charge solution discussed in the previous section. Conversely, if space-charge dominates the bunch length evolution one can expand the formula for large $K_L$ to get
\begin{equation}
    \sigma_{zf} = \frac{2f^2K_L}{\sigma_{z0}^2}
    \label{sclimit}
\end{equation}
which is linear in the bunch charge, but most importantly decreases as $\sigma_{z0}^2$ driving the initial bunch length towards larger values. Note that if by some clever scheme (for example by pre-compensating using an X-band RF cavity as discussed in the last section of the paper) the RF emittance growth in the buncher is eliminated, a longer initial bunch length could be used, reducing space-charge effects and potentially allowing to reach very short bunch lengths.

The optimal initial conditions are thus found as a compromise between the tendency to minimize the emittance growth in the buncher and to reduce the space-charge effects lowering the initial peak current. An estimate for the ideal initial bunch length can be obtained by equating the asymptotic dependencies of the waist size in the space-charge dominated regime and the emittance dominated regime respectively.

For example in the case where the RF emittance is dominated by the quadratic non linearity, we can set Eq. \ref{sclimit} equal to the second expression in Eq. \ref{Eq:simpleformulabunchlength}, and the optimum initial bunch size is given by the solution of the implicit equation: 
\begin{equation}
    \sigma_{z0}^4\approx\frac{\sqrt{2}K_L }{\eta_1|\eta_2|\alpha^3k^3}
    \label{Eq:minimumsigmaz0}
\end{equation}
This result can be substituted into Eq. \ref{Eq:sigmazfwithsc} yielding for the minimum bunch duration at the temporal waist:
\begin{equation}
    \sigma_{zf}=(\sqrt{5}+1)\sqrt{\frac{K_L|\eta_2|}{\sqrt{2}\eta_1^3\alpha k}}
    \label{Eq:minimumsigmazf_sc}
\end{equation}
In the high energy limit ($\gamma_0 \gg 1)$
\begin{equation}
    \sigma_{zf} \sim \sqrt{\frac{3g N f r_c}{\gamma_0^3}}
    \label{Eq:relativistic_sclimit}
\end{equation}

\begin{figure*}[ht]
    \includegraphics[scale=0.55]{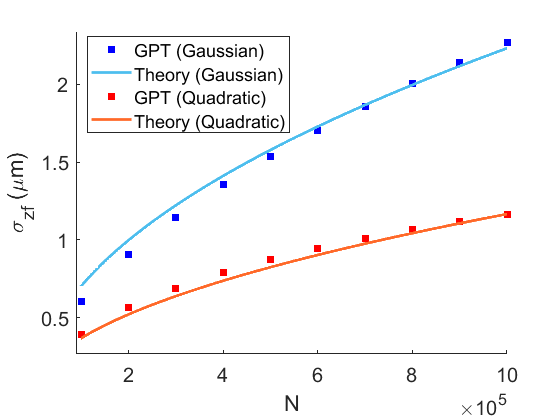}(a)
    \includegraphics[scale=0.55]{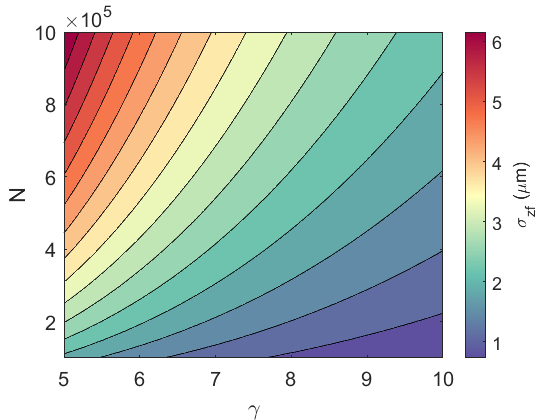}(b)
\caption{
     (a) Minimum bunch length analytical estimates compared with GPT simulation points as the the number of electrons in the beam is varied for the gaussian (blue) and uniformly filled ellipsoidal (orange) distributions (b) Minimum bunch length versus $\gamma$ and $N$ for the Gaussian distribution assuming a constant focal length $f$ = 2 m.}
     \label{fig:varyN}
\end{figure*}

In Fig. \ref{fig:scminagreement}(a) we show the final waist as a function of the injected bunch length obtained from GPT, and overlay Eq. \ref{Eq:sigmazfwithsc} for the case when the total charge is $Q=10^5e = 16~$fC. The dashed black line is the estimate obtained from Eq. \ref{Eq:minimumsigmazf_sc}, which predicts an RMS size of 0.7$\mu$m, in excellent agreement with the GPT prediction at 0.6 $\mu$m. In Fig. \ref{fig:scminagreement}(b) the phase space at the waist for the optimal injection case is shown. The quadratic correlation in phase space is still visible, but an imprint of the space-charge field begins to show up as well. space-charge effects are essentially negligible for large enough initial bunch length. When the initial bunch length is smaller than the optimum, i.e. in the space-charge dominated regime, the size at the waist begins to grow significantly, and unaccounted non-linear space-charge effects begin to play a role. 

Thus far, we have completely neglected the non-linear space-charge field contributions to the envelope evolution. Eventually, these terms will become significant as we consider initially shorter bunches leading to large emittance growth thus disrupting the applicability of the analytical result which has been obtained assuming a constant emittance in the drift. 

This is elucidated in Fig. \ref{fig:scminagreement}(c) and (d), where we show the trace space emittance evolution for the optimum compression case, and the final emittance as a function of input bunch length respectively. In (c), it can be seen that the emittance first increases to the predicted value after the buncher due to the RF non-linearities. After the buncher, the emittance keeps growing in the drift due to space-charge effects. In (d), the emittance calculated from the theory (Eq. \ref{Eq:RFemittance}) is shown along with the final emittance at the waist. The emittance growth induced by space-charge is estimated by subtracting in quadrature the theory from the GPT results. Therein, it can be seen that the optimum input bunch length (60~$\mu$m in this case) occurs at the onset of emittance growth from space-charge. Eventually, as the initial bunch length is further decreased, the emittance growth is dominated by space-charge. 

In Fig. \ref{fig:varyN}(a), we show a comparison of Eq. \ref{Eq:minimumsigmazf_sc} with GPT simulations performed while setting the initial bunch length to Eq. \ref{Eq:minimumsigmaz0} as the charge is varied from $10^5e\to10^6e$. Again, the agreement is excellent. In Fig. \ref{fig:varyN}(b), we utilize Eq. \ref{Eq:minimumsigmazf_sc}, to visualize the dependence on energy and charge when the focal length of the bunching system is set to 2~m.
Note that due to the energy dependence intrinsic in the focal length, as we increase the beam energy, keeping the focal length constant becomes a technological feat 
involving considerations of the breakdown limit in RF cavities as well as available power sources at higher frequencies. 

To go beyond these limits, one would need to reduce the space-charge effects (working to either minimize the geometry factor $g$ in the longitudinal perveance, and/or utilize a charge distribution with more linear self-fields) or minimize the emittance growth in the buncher, for example, pre-compensating for the non-linearities in the input phase space. Both options are in principle feasible and are the subjects of the last two sections of this paper. 

\section{Bunch compression limits for different charge distributions.}
So far we have been assuming a Gaussian temporal profile for the input electron bunch. It is worth investigating the prospect of a more ideal phase space such as the uniformly filled ellipsoid with transverse and longitudinal dimensions $a$ and $z_m$ respectively. An analytic derivation of the bunch length limit can be performed in this case as well. For this, we will adopt the notation of \cite{reiserbook}. The line charge density for this distribution can be written as:
\begin{equation}
    \lambda(z)=\frac{3Q}{4z_m}\left(1-\frac{z^2}{z_m^2}\right)
    \label{Eq.Q}
\end{equation}
where $z_m$ is related to the RMS size of the beam as $z_m=\sqrt{5}\sigma_z$, and likewise $a=\sqrt{5}\sigma_r$. Repeating the calculation from section \ref{sect:rfemittancegrowth} to calculate the induced RF emittance at the exit of the buncher with this line charge density yields: 
\begin{equation}
    \epsilon_{Rf}^2\approx\frac{8}{7}\eta_2^2\alpha^4 k^4 \sigma_{z0}^6
    \label{Eq:Quad_emit}
\end{equation}
similar to what was previously calculated but smaller by a factor $\frac{4}{7}$. This is because, for the same RMS bunch size, the parabolic current profile extends over a smaller interval of RF phases than the long-tailed gaussian distribution, thus reducing the associated RMS emittance growth.

The perveance and geometry parameter of the longitudinal field evolution have been derived for this distribution as well. The geometry parameter is twice the one obtained for the gaussian distribution, i.e., $v_u(A)=2v(A)$. The longitudinal field is given by:
\begin{equation}
    E_z=-\frac{v_u(A)}{4\pi\epsilon_0}\frac{\partial\lambda}{\partial z}
\end{equation}

\begin{figure*}[ht]
    \includegraphics[scale=0.75]{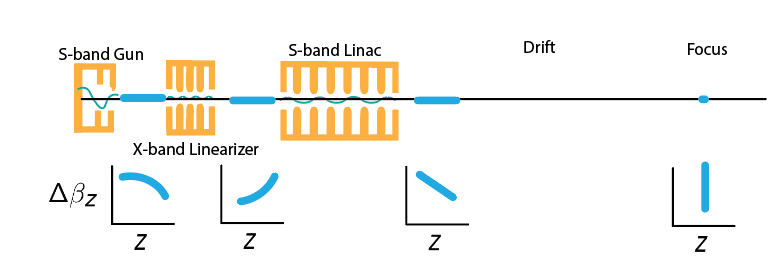}(a)
    \includegraphics[scale=0.365]{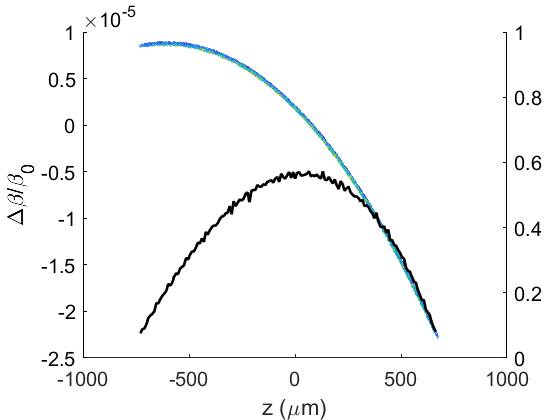}(b)
    \includegraphics[scale=0.365]{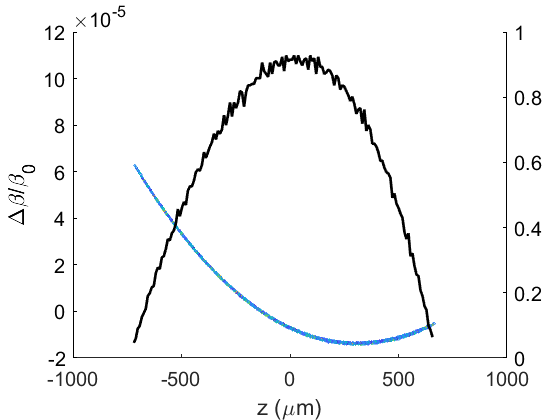}(c)
    \includegraphics[scale=0.365]{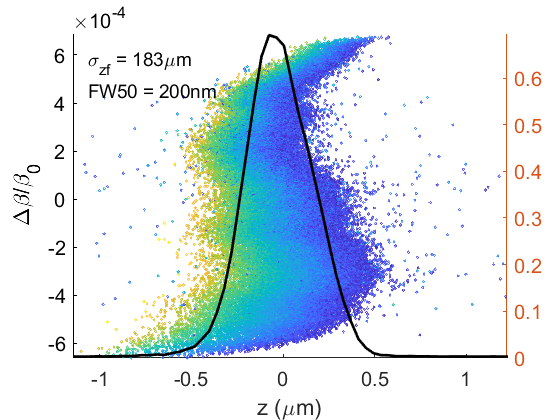}(d)
    \caption{(a) Illustration of beamline setup for the RF emittance growth compensation. A short x-band linearizer is used to compensate for the curvature in $(z,z')$ space imparted by the S-band linac and gun. (b, c, and d) GPT simulation of phase space at the exit of the S-band gun, the exit of the X-band linearizer, and at the longitudinal focus. The temporal current profiles (note the change in the x-axis) are shown in black. The particles are color-coded with their transverse radial coordinate, to highlight that to further improve the bunch compression, the full interplay of transverse and longitudinal dynamics should be considered.} 
    \label{fig:compensation}
\end{figure*}

Then the generalized longitudinal force derived from this field yields the perveance:
\begin{equation}
    K_{L,u}=\frac{g_u(A)N r_c}{\beta^2 \gamma^5}
\end{equation}
where now, $g_u(A)=3v(A)/5\sqrt{5}$. Note, that for the same aspect ratio as the gaussian, the ratio $g_u/g\approx0.33$, indicative of enhanced compression performances. Both the prefactors of the RF emittance growth and the perveance are smaller than those obtained for the Gaussian, so using this distribution, substantial improvements are expected. To this end, we can use this perveance factor and by equating Eq. \ref{sclimit} to Eq. \ref{Eq:min_bunch_length}, while only keeping Eq. \ref{Eq:Quad_emit} as the leading contribution to the emittance, we can again obtain an estimate of the optimum initial bunch length. 

We repeated for this case the benchmarking simulation study by initializing in GPT a uniformly filled ellipsoidal distribution with optimal input bunch length and comparing the results with the analytical predictions as a function of beam charge. The results of the scan are shown in Fig. \ref{fig:varyN} (a). Compared to the gaussian current profile, this distribution allows nearly an overall factor of two improvements in optimum bunch length, consistent with the expectations. 

\section{X-band cavity compensation}

Another option that we consider to obtain even shorter bunch lengths is the utilization of a second higher frequency cavity to compensate for the curvature imparted on the longitudinal phase space by the RF buncher fields and minimize the emittance growth. 

 \begin{table}[hb]
   \centering
   \caption{Parameters of simluation.}
   \begin{tabular}{lcc}
       \toprule
       \textbf{Parameter} & \textbf{}                      & \textbf{Value} \\
Charge & & $10^5$e\\
Laser Spot Size & & 40 $\mu$m\\
Cathode MTE & & 0.5 eV\\
Laser pulse length & & 1.65 ps (rms) \\
S-band Gun Gradient & & 100 MV/m\\
S-band Gun Phase & &  26.57 deg\\
Gun Exit Beam Energy & &  4.6 MeV\\
X-band Linearizer Accelerating Gradient & & 20 MV/m\\
X-band Linearizer Length & & 0.109m\\
X-band Linearizer Phase  (off-crest) & &  180 deg\\
S-band Linac Accelerating Gradient & & 14MV/m\\
S-band Linac Accelerating Length & & 0.625m\\
S-band Linac Phase (off-crest) & &  -90 deg\\
S-band Linac Exit Beam Energy & &  3.4 MeV\\
   \end{tabular}
   \label{parameters2}
\end{table}

Before analyzing this case in detail, we note that our discussion so far has assumed an initially flat and uncorrelated longitudinal phase space. In a realistic system, the beam out of the gun would typically present already some correlations (often of non-linear nature) between energy and time along the bunch. Here we use a simulation starting from the cathode to allow a detailed study of the compensation with a realistic input longitudinal phase space.

For this section, the setup is closely mimicking the configuration of the UCLA Pegasus beamline \cite{alesini2015new} consisting of a 1.6 Cell S-band (2.856 GHz) gun, followed by an X-band (9.6 GHz) cavity situated at 1.1 m from the cathode for longitudinal emittance compensation, then following that is an S-band (2.856 GHz) 11 cell linac operating as a buncher. The buncher entrance is 1.4m downstream of the cathode. An illustration of the beamline is shown in Fig. \ref{fig:compensation}(a). 

In practice, the optimization begins by initially finding the amplitude and phase of the X-band compensation cavity to cancel the quadratic curvature at the exit of the RF gun. We then further increase the amplitude past that point to also pre-compensate the quadratic non-linearity associated with the relativistic transport, which in the relativistic case is the leading contribution to the emittance growth in the $(z,z')$ trace space. 

In Table. \ref{parameters2}, the nominal parameters at the optimal point are given, namely amplitudes, lengths, and phases of the respective cavities, along with initial beam parameters. The initial pulse profile has a quadratic shape. A cartoon of the phase space evolution in successive drifts is shown as well, illustrating the behavior of the predicted behavior of the curvature at each stage of the scheme. In Fig. \ref{fig:compensation}(b-d), the phase spaces after the gun, X-band, and at the longitudinal waist plane are shown respectively. Note the reversal of concavity of the longitudinal phase space from Fig. \ref{fig:compensation}(b) to Fig. \ref{fig:compensation}(c) caused by the deceleration in the X-band cavity. The concavity of the phase space before the buncher is set to pre-cancel the additional curvature imparted by the relativistic energy-velocity relation. 

To understand the quality of this numerical optimization in the context of the analytical formulas discussed earlier in the paper, we extract the uncorrelated energy spread at the exit of the buncher by calculating the RMS of the distribution after subtracting the linear chirp. If the non-linear emittance growth is compensated, we can expect a bunch length limited only by the uncorrelated energy spread. Equation \ref{Eq:simpleformulabunchlength} evaluated with an RMS uncorrelated energy spread of $10^{-5}$, final beam energy  $\gamma=7.65$, and focal length of 0.8 m as extracted from simulation, yield an estimate for the ultimate limit in bunch length of 170nm or 510 as. This value is very close to the result of the optimized simulation shown in Fig. \ref{fig:compensation}(d) where the waist size is 183nm or 600 as. Looking at the phase space distribution we can see that the bunch length at the waist is mainly limited by uncorrelated energy spread. Color coding the particle with their radial position indicates that a significant contribution in this energy spread is the residual transverse correlations. In principle removing those would allow even shorter bunch lengths.

This scheme does provide a direct path to sub-fs bunch length, but the synchronization of the drive signals for two different resonant frequency RF cavities is a significant technological challenge that has to be addressed before moving forward to implement this in a user-facility UED beamline. The case is discussed here mainly as an example of the insights offered by the analytical framework that we developed, which suggests that properly shaping the drive laser and compensating the emittance growth does provide a path towards much shorter bunch lengths. 


\section{Conclusion}

In conclusion, in this paper, we have developed an analytical framework that yields a simple estimate for the minimum bunch length achievable in an RF compression beamline. Besides UED, the formulas in this paper might be useful in the optimization of RF compression for other ultrashort electron beam applications including radiation generation and injection into very high-frequency THz-driven \cite{nanniterahertz} and laser-driven accelerators \cite{cesar:dla}. 

The formalism of the envelope equation allows us to include in the analysis various competing effects in the dynamics, including the longitudinal emittance growth due to the RF curvature, the relativistic beam transport, and the space-charge forces. The results indicate an optimum initial bunch length condition (which can be satisfied by adjusting the laser pulse length on the cathode for example) exists which compromises the RF-induced emittance growth and the effects of the longitudinal self-fields. The simplicity of the reported expressions mainly stems from the fact that we approximated the effect of the interplay of the transverse and longitudinal dynamics with a simple constant order-of-unity geometry factor in the longitudinal perveance. We have also limited the expansion yielding the non-linear terms in the emittance growth to second and third-order, but in principle, all the higher-order terms can be calculated (and compensated if enough independent knobs are added/available on the beamline). 
While the initial conditions assumed in the derivations are an initially unchirped longitudinal phase space and realistic beams out of photoinjectors do typically present more complex phase-space distributions, the results obtained here provide useful estimates of the compression limits in a given configuration which can be used as a starting point for numerical optimizations. The scaling laws will be helpful to guide parameter choices in the design of new setups. Most importantly, these results highlight the main contributions to the final bunch length and suggest possible paths to follow to further improve the compression and achieve sub-fs bunch lengths. 

\begin{acknowledgments}
This work has been partially funded by the National Science Foundation (NSF) (PHY-1734215); National Science Foundation under the STROBE Science and Technology Center Grant No. DMR-1548924
and the MRI 
The authors want to acknowledge D. Filippetto, A. Kogar and J. Maxson for insightful discussions leading to the calculations in this paper over the years. 
\end{acknowledgments}

\bibliography{RF_UED}

\providecommand{\noopsort}[1]{}\providecommand{\singleletter}[1]{#1}%
\begin{thebibliography}{21}%
\makeatletter
\providecommand \@ifxundefined [1]{%
 \@ifx{#1\undefined}
}%
\providecommand \@ifnum [1]{%
 \ifnum #1\expandafter \@firstoftwo
 \else \expandafter \@secondoftwo
 \fi
}%
\providecommand \@ifx [1]{%
 \ifx #1\expandafter \@firstoftwo
 \else \expandafter \@secondoftwo
 \fi
}%
\providecommand \natexlab [1]{#1}%
\providecommand \enquote  [1]{``#1''}%
\providecommand \bibnamefont  [1]{#1}%
\providecommand \bibfnamefont [1]{#1}%
\providecommand \citenamefont [1]{#1}%
\providecommand \href@noop [0]{\@secondoftwo}%
\providecommand \href [0]{\begingroup \@sanitize@url \@href}%
\providecommand \@href[1]{\@@startlink{#1}\@@href}%
\providecommand \@@href[1]{\endgroup#1\@@endlink}%
\providecommand \@sanitize@url [0]{\catcode `\\12\catcode `\$12\catcode
  `\&12\catcode `\#12\catcode `\^12\catcode `\_12\catcode `\%12\relax}%
\providecommand \@@startlink[1]{}%
\providecommand \@@endlink[0]{}%
\providecommand \url  [0]{\begingroup\@sanitize@url \@url }%
\providecommand \@url [1]{\endgroup\@href {#1}{\urlprefix }}%
\providecommand \urlprefix  [0]{URL }%
\providecommand \Eprint [0]{\href }%
\providecommand \doibase [0]{https://doi.org/}%
\providecommand \selectlanguage [0]{\@gobble}%
\providecommand \bibinfo  [0]{\@secondoftwo}%
\providecommand \bibfield  [0]{\@secondoftwo}%
\providecommand \translation [1]{[#1]}%
\providecommand \BibitemOpen [0]{}%
\providecommand \bibitemStop [0]{}%
\providecommand \bibitemNoStop [0]{.\EOS\space}%
\providecommand \EOS [0]{\spacefactor3000\relax}%
\providecommand \BibitemShut  [1]{\csname bibitem#1\endcsname}%
\let\auto@bib@innerbib\@empty
\bibitem [{\citenamefont {Zewail}(2006)}]{zewail}%
  \BibitemOpen
  \bibfield  {author} {\bibinfo {author} {\bibfnamefont {A.~H.}\ \bibnamefont
  {Zewail}},\ }\bibfield  {title} {\bibinfo {title} {4d ultrafast electron
  diffraction, crystallography, and microscopy},\ }\href@noop {} {\bibfield
  {journal} {\bibinfo  {journal} {Annu. Rev. Phys. Chem.}\ }\textbf {\bibinfo
  {volume} {57}},\ \bibinfo {pages} {65} (\bibinfo {year} {2006})}\BibitemShut
  {NoStop}%
\bibitem [{\citenamefont {Sciaini}\ and\ \citenamefont
  {Miller}(2011)}]{sciainiMiller}%
  \BibitemOpen
  \bibfield  {author} {\bibinfo {author} {\bibfnamefont {G.}~\bibnamefont
  {Sciaini}}\ and\ \bibinfo {author} {\bibfnamefont {R.~D.}\ \bibnamefont
  {Miller}},\ }\bibfield  {title} {\bibinfo {title} {Femtosecond electron
  diffraction: heralding the era of atomically resolved dynamics},\ }\href@noop
  {} {\bibfield  {journal} {\bibinfo  {journal} {Reports on Progress in
  Physics}\ }\textbf {\bibinfo {volume} {74}},\ \bibinfo {pages} {096101}
  (\bibinfo {year} {2011})}\BibitemShut {NoStop}%
\bibitem [{\citenamefont {Siwick}\ \emph {et~al.}(2003)\citenamefont {Siwick},
  \citenamefont {Dwyer}, \citenamefont {Jordan},\ and\ \citenamefont
  {Miller}}]{siwick2003}%
  \BibitemOpen
  \bibfield  {author} {\bibinfo {author} {\bibfnamefont {B.~J.}\ \bibnamefont
  {Siwick}}, \bibinfo {author} {\bibfnamefont {J.~R.}\ \bibnamefont {Dwyer}},
  \bibinfo {author} {\bibfnamefont {R.~E.}\ \bibnamefont {Jordan}},\ and\
  \bibinfo {author} {\bibfnamefont {R.~D.}\ \bibnamefont {Miller}},\ }\bibfield
   {title} {\bibinfo {title} {An atomic-level view of melting using femtosecond
  electron diffraction},\ }\href@noop {} {\bibfield  {journal} {\bibinfo
  {journal} {Science}\ }\textbf {\bibinfo {volume} {302}},\ \bibinfo {pages}
  {1382} (\bibinfo {year} {2003})}\BibitemShut {NoStop}%
\bibitem [{\citenamefont {Weathersby}\ \emph {et~al.}(2015)\citenamefont
  {Weathersby}, \citenamefont {Brown}, \citenamefont {Centurion}, \citenamefont
  {Chase}, \citenamefont {Coffee}, \citenamefont {Corbett}, \citenamefont
  {Eichner}, \citenamefont {Frisch}, \citenamefont {Fry}, \citenamefont
  {G{\"u}hr} \emph {et~al.}}]{weathersby2015mega}%
  \BibitemOpen
  \bibfield  {author} {\bibinfo {author} {\bibfnamefont {S.}~\bibnamefont
  {Weathersby}}, \bibinfo {author} {\bibfnamefont {G.}~\bibnamefont {Brown}},
  \bibinfo {author} {\bibfnamefont {M.}~\bibnamefont {Centurion}}, \bibinfo
  {author} {\bibfnamefont {T.}~\bibnamefont {Chase}}, \bibinfo {author}
  {\bibfnamefont {R.}~\bibnamefont {Coffee}}, \bibinfo {author} {\bibfnamefont
  {J.}~\bibnamefont {Corbett}}, \bibinfo {author} {\bibfnamefont
  {J.}~\bibnamefont {Eichner}}, \bibinfo {author} {\bibfnamefont
  {J.}~\bibnamefont {Frisch}}, \bibinfo {author} {\bibfnamefont
  {A.}~\bibnamefont {Fry}}, \bibinfo {author} {\bibfnamefont {M.}~\bibnamefont
  {G{\"u}hr}}, \emph {et~al.},\ }\bibfield  {title} {\bibinfo {title}
  {Mega-electron-volt ultrafast electron diffraction at slac national
  accelerator laboratory},\ }\href@noop {} {\bibfield  {journal} {\bibinfo
  {journal} {Review of Scientific Instruments}\ }\textbf {\bibinfo {volume}
  {86}},\ \bibinfo {pages} {073702} (\bibinfo {year} {2015})}\BibitemShut
  {NoStop}%
\bibitem [{\citenamefont {Zhu}\ \emph {et~al.}(2015)\citenamefont {Zhu},
  \citenamefont {Zhu}, \citenamefont {Hidaka}, \citenamefont {Wu},
  \citenamefont {Cao}, \citenamefont {Berger}, \citenamefont {Geck},
  \citenamefont {Kraus}, \citenamefont {Pjerov}, \citenamefont {Shen} \emph
  {et~al.}}]{zhu2015femtosecond}%
  \BibitemOpen
  \bibfield  {author} {\bibinfo {author} {\bibfnamefont {P.}~\bibnamefont
  {Zhu}}, \bibinfo {author} {\bibfnamefont {Y.}~\bibnamefont {Zhu}}, \bibinfo
  {author} {\bibfnamefont {Y.}~\bibnamefont {Hidaka}}, \bibinfo {author}
  {\bibfnamefont {L.}~\bibnamefont {Wu}}, \bibinfo {author} {\bibfnamefont
  {J.}~\bibnamefont {Cao}}, \bibinfo {author} {\bibfnamefont {H.}~\bibnamefont
  {Berger}}, \bibinfo {author} {\bibfnamefont {J.}~\bibnamefont {Geck}},
  \bibinfo {author} {\bibfnamefont {R.}~\bibnamefont {Kraus}}, \bibinfo
  {author} {\bibfnamefont {S.}~\bibnamefont {Pjerov}}, \bibinfo {author}
  {\bibfnamefont {Y.}~\bibnamefont {Shen}}, \emph {et~al.},\ }\bibfield
  {title} {\bibinfo {title} {Femtosecond time-resolved mev electron
  diffraction},\ }\href@noop {} {\bibfield  {journal} {\bibinfo  {journal} {New
  Journal of Physics}\ }\textbf {\bibinfo {volume} {17}},\ \bibinfo {pages}
  {063004} (\bibinfo {year} {2015})}\BibitemShut {NoStop}%
\bibitem [{\citenamefont {Manz}\ \emph {et~al.}(2015)\citenamefont {Manz},
  \citenamefont {Casandruc}, \citenamefont {Zhang}, \citenamefont {Zhong},
  \citenamefont {Loch}, \citenamefont {Marx}, \citenamefont {Hasegawa},
  \citenamefont {Liu}, \citenamefont {Bayesteh}, \citenamefont {Delsim-Hashemi}
  \emph {et~al.}}]{manz2015mapping}%
  \BibitemOpen
  \bibfield  {author} {\bibinfo {author} {\bibfnamefont {S.}~\bibnamefont
  {Manz}}, \bibinfo {author} {\bibfnamefont {A.}~\bibnamefont {Casandruc}},
  \bibinfo {author} {\bibfnamefont {D.}~\bibnamefont {Zhang}}, \bibinfo
  {author} {\bibfnamefont {Y.}~\bibnamefont {Zhong}}, \bibinfo {author}
  {\bibfnamefont {R.~A.}\ \bibnamefont {Loch}}, \bibinfo {author}
  {\bibfnamefont {A.}~\bibnamefont {Marx}}, \bibinfo {author} {\bibfnamefont
  {T.}~\bibnamefont {Hasegawa}}, \bibinfo {author} {\bibfnamefont {L.~C.}\
  \bibnamefont {Liu}}, \bibinfo {author} {\bibfnamefont {S.}~\bibnamefont
  {Bayesteh}}, \bibinfo {author} {\bibfnamefont {H.}~\bibnamefont
  {Delsim-Hashemi}}, \emph {et~al.},\ }\bibfield  {title} {\bibinfo {title}
  {Mapping atomic motions with ultrabright electrons: towards fundamental
  limits in space-time resolution},\ }\href@noop {} {\bibfield  {journal}
  {\bibinfo  {journal} {Faraday discussions}\ }\textbf {\bibinfo {volume}
  {177}},\ \bibinfo {pages} {467} (\bibinfo {year} {2015})}\BibitemShut
  {NoStop}%
\bibitem [{\citenamefont {Filippetto}\ and\ \citenamefont
  {Qian}(2016)}]{filippetto2016design}%
  \BibitemOpen
  \bibfield  {author} {\bibinfo {author} {\bibfnamefont {D.}~\bibnamefont
  {Filippetto}}\ and\ \bibinfo {author} {\bibfnamefont {H.}~\bibnamefont
  {Qian}},\ }\bibfield  {title} {\bibinfo {title} {Design of a high-flux
  instrument for ultrafast electron diffraction and microscopy},\ }\href@noop
  {} {\bibfield  {journal} {\bibinfo  {journal} {Journal of Physics B: Atomic,
  Molecular and Optical Physics}\ }\textbf {\bibinfo {volume} {49}},\ \bibinfo
  {pages} {104003} (\bibinfo {year} {2016})}\BibitemShut {NoStop}%
\bibitem [{\citenamefont {Van~Oudheusden}\ \emph {et~al.}(2007)\citenamefont
  {Van~Oudheusden}, \citenamefont {De~Jong}, \citenamefont {Van~der Geer},
  \citenamefont {’t Root}, \citenamefont {Luiten},\ and\ \citenamefont
  {Siwick}}]{van2007electron}%
  \BibitemOpen
  \bibfield  {author} {\bibinfo {author} {\bibfnamefont {T.}~\bibnamefont
  {Van~Oudheusden}}, \bibinfo {author} {\bibfnamefont {E.}~\bibnamefont
  {De~Jong}}, \bibinfo {author} {\bibfnamefont {S.}~\bibnamefont {Van~der
  Geer}}, \bibinfo {author} {\bibfnamefont {W.~O.}\ \bibnamefont {’t Root}},
  \bibinfo {author} {\bibfnamefont {O.}~\bibnamefont {Luiten}},\ and\ \bibinfo
  {author} {\bibfnamefont {B.}~\bibnamefont {Siwick}},\ }\bibfield  {title}
  {\bibinfo {title} {Electron source concept for single-shot sub-100 fs
  electron diffraction in the 100 kev range},\ }\href@noop {} {\bibfield
  {journal} {\bibinfo  {journal} {Journal of Applied Physics}\ }\textbf
  {\bibinfo {volume} {102}},\ \bibinfo {pages} {093501} (\bibinfo {year}
  {2007})}\BibitemShut {NoStop}%
\bibitem [{\citenamefont {Van~Oudheusden}\ \emph {et~al.}(2010)\citenamefont
  {Van~Oudheusden}, \citenamefont {Pasmans}, \citenamefont {Van Der~Geer},
  \citenamefont {De~Loos}, \citenamefont {Van Der~Wiel},\ and\ \citenamefont
  {Luiten}}]{van2010compression}%
  \BibitemOpen
  \bibfield  {author} {\bibinfo {author} {\bibfnamefont {T.}~\bibnamefont
  {Van~Oudheusden}}, \bibinfo {author} {\bibfnamefont {P.}~\bibnamefont
  {Pasmans}}, \bibinfo {author} {\bibfnamefont {S.}~\bibnamefont {Van
  Der~Geer}}, \bibinfo {author} {\bibfnamefont {M.}~\bibnamefont {De~Loos}},
  \bibinfo {author} {\bibfnamefont {M.}~\bibnamefont {Van Der~Wiel}},\ and\
  \bibinfo {author} {\bibfnamefont {O.}~\bibnamefont {Luiten}},\ }\bibfield
  {title} {\bibinfo {title} {Compression of subrelativistic
  space-charge-dominated electron bunches for single-shot femtosecond electron
  diffraction},\ }\href@noop {} {\bibfield  {journal} {\bibinfo  {journal}
  {Physical review letters}\ }\textbf {\bibinfo {volume} {105}},\ \bibinfo
  {pages} {264801} (\bibinfo {year} {2010})}\BibitemShut {NoStop}%
\bibitem [{\citenamefont {Maxson}\ \emph {et~al.}(2017)\citenamefont {Maxson},
  \citenamefont {Cesar}, \citenamefont {Calmasini}, \citenamefont {Ody},
  \citenamefont {Musumeci},\ and\ \citenamefont {Alesini}}]{maxson2017}%
  \BibitemOpen
  \bibfield  {author} {\bibinfo {author} {\bibfnamefont {J.}~\bibnamefont
  {Maxson}}, \bibinfo {author} {\bibfnamefont {D.}~\bibnamefont {Cesar}},
  \bibinfo {author} {\bibfnamefont {G.}~\bibnamefont {Calmasini}}, \bibinfo
  {author} {\bibfnamefont {A.}~\bibnamefont {Ody}}, \bibinfo {author}
  {\bibfnamefont {P.}~\bibnamefont {Musumeci}},\ and\ \bibinfo {author}
  {\bibfnamefont {D.}~\bibnamefont {Alesini}},\ }\bibfield  {title} {\bibinfo
  {title} {Direct measurement of sub-10 fs relativistic electron beams with
  ultralow emittance},\ }\href@noop {} {\bibfield  {journal} {\bibinfo
  {journal} {Physical review letters}\ }\textbf {\bibinfo {volume} {118}},\
  \bibinfo {pages} {154802} (\bibinfo {year} {2017})}\BibitemShut {NoStop}%
\bibitem [{\citenamefont {Zhao}\ \emph {et~al.}(2018)\citenamefont {Zhao},
  \citenamefont {Wang}, \citenamefont {Lu}, \citenamefont {Wang}, \citenamefont
  {Hu}, \citenamefont {Wang}, \citenamefont {Qi}, \citenamefont {Jiang},
  \citenamefont {Liu}, \citenamefont {Ma} \emph {et~al.}}]{zhao2018terahertz}%
  \BibitemOpen
  \bibfield  {author} {\bibinfo {author} {\bibfnamefont {L.}~\bibnamefont
  {Zhao}}, \bibinfo {author} {\bibfnamefont {Z.}~\bibnamefont {Wang}}, \bibinfo
  {author} {\bibfnamefont {C.}~\bibnamefont {Lu}}, \bibinfo {author}
  {\bibfnamefont {R.}~\bibnamefont {Wang}}, \bibinfo {author} {\bibfnamefont
  {C.}~\bibnamefont {Hu}}, \bibinfo {author} {\bibfnamefont {P.}~\bibnamefont
  {Wang}}, \bibinfo {author} {\bibfnamefont {J.}~\bibnamefont {Qi}}, \bibinfo
  {author} {\bibfnamefont {T.}~\bibnamefont {Jiang}}, \bibinfo {author}
  {\bibfnamefont {S.}~\bibnamefont {Liu}}, \bibinfo {author} {\bibfnamefont
  {Z.}~\bibnamefont {Ma}}, \emph {et~al.},\ }\bibfield  {title} {\bibinfo
  {title} {Terahertz streaking of few-femtosecond relativistic electron
  beams},\ }\href@noop {} {\bibfield  {journal} {\bibinfo  {journal} {Physical
  Review X}\ }\textbf {\bibinfo {volume} {8}},\ \bibinfo {pages} {021061}
  (\bibinfo {year} {2018})}\BibitemShut {NoStop}%
\bibitem [{\citenamefont {Snively}\ \emph {et~al.}(2020)\citenamefont
  {Snively}, \citenamefont {Othman}, \citenamefont {Kozina}, \citenamefont
  {Ofori-Okai}, \citenamefont {Weathersby}, \citenamefont {Park}, \citenamefont
  {Shen}, \citenamefont {Wang}, \citenamefont {Hoffmann}, \citenamefont {Li}
  \emph {et~al.}}]{snively2020femtosecond}%
  \BibitemOpen
  \bibfield  {author} {\bibinfo {author} {\bibfnamefont {E.}~\bibnamefont
  {Snively}}, \bibinfo {author} {\bibfnamefont {M.}~\bibnamefont {Othman}},
  \bibinfo {author} {\bibfnamefont {M.}~\bibnamefont {Kozina}}, \bibinfo
  {author} {\bibfnamefont {B.}~\bibnamefont {Ofori-Okai}}, \bibinfo {author}
  {\bibfnamefont {S.}~\bibnamefont {Weathersby}}, \bibinfo {author}
  {\bibfnamefont {S.}~\bibnamefont {Park}}, \bibinfo {author} {\bibfnamefont
  {X.}~\bibnamefont {Shen}}, \bibinfo {author} {\bibfnamefont {X.}~\bibnamefont
  {Wang}}, \bibinfo {author} {\bibfnamefont {M.}~\bibnamefont {Hoffmann}},
  \bibinfo {author} {\bibfnamefont {R.}~\bibnamefont {Li}}, \emph {et~al.},\
  }\bibfield  {title} {\bibinfo {title} {Femtosecond compression dynamics and
  timing jitter suppression in a thz-driven electron bunch compressor},\
  }\href@noop {} {\bibfield  {journal} {\bibinfo  {journal} {Physical review
  letters}\ }\textbf {\bibinfo {volume} {124}},\ \bibinfo {pages} {054801}
  (\bibinfo {year} {2020})}\BibitemShut {NoStop}%
\bibitem [{\citenamefont {Kealhofer}\ \emph {et~al.}(2016)\citenamefont
  {Kealhofer}, \citenamefont {Schneider}, \citenamefont {Ehberger},
  \citenamefont {Ryabov}, \citenamefont {Krausz},\ and\ \citenamefont
  {Baum}}]{kealhofer2016all}%
  \BibitemOpen
  \bibfield  {author} {\bibinfo {author} {\bibfnamefont {C.}~\bibnamefont
  {Kealhofer}}, \bibinfo {author} {\bibfnamefont {W.}~\bibnamefont
  {Schneider}}, \bibinfo {author} {\bibfnamefont {D.}~\bibnamefont {Ehberger}},
  \bibinfo {author} {\bibfnamefont {A.}~\bibnamefont {Ryabov}}, \bibinfo
  {author} {\bibfnamefont {F.}~\bibnamefont {Krausz}},\ and\ \bibinfo {author}
  {\bibfnamefont {P.}~\bibnamefont {Baum}},\ }\bibfield  {title} {\bibinfo
  {title} {All-optical control and metrology of electron pulses},\ }\href@noop
  {} {\bibfield  {journal} {\bibinfo  {journal} {Science}\ }\textbf {\bibinfo
  {volume} {352}},\ \bibinfo {pages} {429} (\bibinfo {year}
  {2016})}\BibitemShut {NoStop}%
\bibitem [{\citenamefont {De~Loos}\ and\ \citenamefont {Van~der
  Geer}(1996)}]{GPT}%
  \BibitemOpen
  \bibfield  {author} {\bibinfo {author} {\bibfnamefont {M.}~\bibnamefont
  {De~Loos}}\ and\ \bibinfo {author} {\bibfnamefont {S.}~\bibnamefont {Van~der
  Geer}},\ }\bibfield  {title} {\bibinfo {title} {General particle tracer: A
  new 3d code for accelerator and beamline design},\ }in\ \href@noop {} {\emph
  {\bibinfo {booktitle} {5th European Particle Accelerator Conference}}}\
  (\bibinfo {year} {1996})\ p.\ \bibinfo {pages} {1241}\BibitemShut {NoStop}%
\bibitem [{\citenamefont {Floettmann}(2014)}]{floettmann:subfs}%
  \BibitemOpen
  \bibfield  {author} {\bibinfo {author} {\bibfnamefont {K.}~\bibnamefont
  {Floettmann}},\ }\bibfield  {title} {\bibinfo {title} {Generation of sub-fs
  electron beams at few-mev energies},\ }\href@noop {} {\bibfield  {journal}
  {\bibinfo  {journal} {Nuclear Instruments and Methods in Physics Research
  Section A: Accelerators, Spectrometers, Detectors and Associated Equipment}\
  }\textbf {\bibinfo {volume} {740}},\ \bibinfo {pages} {34} (\bibinfo {year}
  {2014})}\BibitemShut {NoStop}%
\bibitem [{\citenamefont {Reiser}\ and\ \citenamefont
  {O'Shea}(1994)}]{reiserbook}%
  \BibitemOpen
  \bibfield  {author} {\bibinfo {author} {\bibfnamefont {M.}~\bibnamefont
  {Reiser}}\ and\ \bibinfo {author} {\bibfnamefont {P.}~\bibnamefont
  {O'Shea}},\ }\href@noop {} {\emph {\bibinfo {title} {Theory and design of
  charged particle beams}}},\ Vol.\ \bibinfo {volume} {312}\ (\bibinfo
  {publisher} {Wiley Online Library},\ \bibinfo {year} {1994})\BibitemShut
  {NoStop}%
\bibitem [{\citenamefont {Wiedemann}(2015)}]{wiedemann}%
  \BibitemOpen
  \bibfield  {author} {\bibinfo {author} {\bibfnamefont {H.}~\bibnamefont
  {Wiedemann}},\ }\href@noop {} {\emph {\bibinfo {title} {Particle accelerator
  physics}}}\ (\bibinfo  {publisher} {Springer Nature},\ \bibinfo {year}
  {2015})\BibitemShut {NoStop}%
\bibitem [{\citenamefont {Zeitler}\ \emph {et~al.}(2015)\citenamefont
  {Zeitler}, \citenamefont {Floettmann},\ and\ \citenamefont
  {Gr{\"u}ner}}]{zeitler}%
  \BibitemOpen
  \bibfield  {author} {\bibinfo {author} {\bibfnamefont {B.}~\bibnamefont
  {Zeitler}}, \bibinfo {author} {\bibfnamefont {K.}~\bibnamefont
  {Floettmann}},\ and\ \bibinfo {author} {\bibfnamefont {F.}~\bibnamefont
  {Gr{\"u}ner}},\ }\bibfield  {title} {\bibinfo {title} {Linearization of the
  longitudinal phase space without higher harmonic field},\ }\href@noop {}
  {\bibfield  {journal} {\bibinfo  {journal} {Physical Review Special
  Topics-Accelerators and Beams}\ }\textbf {\bibinfo {volume} {18}},\ \bibinfo
  {pages} {120102} (\bibinfo {year} {2015})}\BibitemShut {NoStop}%
\bibitem [{\citenamefont {Alesini}\ \emph {et~al.}(2015)\citenamefont
  {Alesini}, \citenamefont {Battisti}, \citenamefont {Ferrario}, \citenamefont
  {Foggetta}, \citenamefont {Lollo}, \citenamefont {Ficcadenti}, \citenamefont
  {Pettinacci}, \citenamefont {Custodio}, \citenamefont {Pirez}, \citenamefont
  {Musumeci} \emph {et~al.}}]{alesini2015new}%
  \BibitemOpen
  \bibfield  {author} {\bibinfo {author} {\bibfnamefont {D.}~\bibnamefont
  {Alesini}}, \bibinfo {author} {\bibfnamefont {A.}~\bibnamefont {Battisti}},
  \bibinfo {author} {\bibfnamefont {M.}~\bibnamefont {Ferrario}}, \bibinfo
  {author} {\bibfnamefont {L.}~\bibnamefont {Foggetta}}, \bibinfo {author}
  {\bibfnamefont {V.}~\bibnamefont {Lollo}}, \bibinfo {author} {\bibfnamefont
  {L.}~\bibnamefont {Ficcadenti}}, \bibinfo {author} {\bibfnamefont
  {V.}~\bibnamefont {Pettinacci}}, \bibinfo {author} {\bibfnamefont
  {S.}~\bibnamefont {Custodio}}, \bibinfo {author} {\bibfnamefont
  {E.}~\bibnamefont {Pirez}}, \bibinfo {author} {\bibfnamefont
  {P.}~\bibnamefont {Musumeci}}, \emph {et~al.},\ }\bibfield  {title} {\bibinfo
  {title} {New technology based on clamping for high gradient radio frequency
  photogun},\ }\href@noop {} {\bibfield  {journal} {\bibinfo  {journal}
  {Physical Review Special Topics-Accelerators and Beams}\ }\textbf {\bibinfo
  {volume} {18}},\ \bibinfo {pages} {092001} (\bibinfo {year}
  {2015})}\BibitemShut {NoStop}%
\bibitem [{\citenamefont {Nanni}\ \emph {et~al.}(2015)\citenamefont {Nanni},
  \citenamefont {Huang}, \citenamefont {Hong}, \citenamefont {Ravi},
  \citenamefont {Fallahi}, \citenamefont {Moriena}, \citenamefont {Miller},\
  and\ \citenamefont {K{\"a}rtner}}]{nanniterahertz}%
  \BibitemOpen
  \bibfield  {author} {\bibinfo {author} {\bibfnamefont {E.~A.}\ \bibnamefont
  {Nanni}}, \bibinfo {author} {\bibfnamefont {W.~R.}\ \bibnamefont {Huang}},
  \bibinfo {author} {\bibfnamefont {K.-H.}\ \bibnamefont {Hong}}, \bibinfo
  {author} {\bibfnamefont {K.}~\bibnamefont {Ravi}}, \bibinfo {author}
  {\bibfnamefont {A.}~\bibnamefont {Fallahi}}, \bibinfo {author} {\bibfnamefont
  {G.}~\bibnamefont {Moriena}}, \bibinfo {author} {\bibfnamefont {R.~D.}\
  \bibnamefont {Miller}},\ and\ \bibinfo {author} {\bibfnamefont {F.~X.}\
  \bibnamefont {K{\"a}rtner}},\ }\bibfield  {title} {\bibinfo {title}
  {Terahertz-driven linear electron acceleration},\ }\href@noop {} {\bibfield
  {journal} {\bibinfo  {journal} {Nature communications}\ }\textbf {\bibinfo
  {volume} {6}},\ \bibinfo {pages} {1} (\bibinfo {year} {2015})}\BibitemShut
  {NoStop}%
\bibitem [{\citenamefont {Cesar}\ \emph {et~al.}(2018)\citenamefont {Cesar},
  \citenamefont {Maxson}, \citenamefont {Shen}, \citenamefont {Wootton},
  \citenamefont {Tan}, \citenamefont {England},\ and\ \citenamefont
  {Musumeci}}]{cesar:dla}%
  \BibitemOpen
  \bibfield  {author} {\bibinfo {author} {\bibfnamefont {D.}~\bibnamefont
  {Cesar}}, \bibinfo {author} {\bibfnamefont {J.}~\bibnamefont {Maxson}},
  \bibinfo {author} {\bibfnamefont {X.}~\bibnamefont {Shen}}, \bibinfo {author}
  {\bibfnamefont {K.}~\bibnamefont {Wootton}}, \bibinfo {author} {\bibfnamefont
  {S.}~\bibnamefont {Tan}}, \bibinfo {author} {\bibfnamefont {R.}~\bibnamefont
  {England}},\ and\ \bibinfo {author} {\bibfnamefont {P.}~\bibnamefont
  {Musumeci}},\ }\bibfield  {title} {\bibinfo {title} {Enhanced energy gain in
  a dielectric laser accelerator using a tilted pulse front laser},\
  }\href@noop {} {\bibfield  {journal} {\bibinfo  {journal} {Optics express}\
  }\textbf {\bibinfo {volume} {26}},\ \bibinfo {pages} {29216} (\bibinfo {year}
  {2018})}\BibitemShut {NoStop}%
\end{thebibliography}%


\newpage
\clearpage

\newpage
\clearpage

\end{document}